\documentclass[twocolumn,showpacs,preprintnumbers,eqsecnum]{revtex4}
\usepackage{graphicx,dcolumn,bm,amsmath,amssymb}

\begin{document}
\title{\ \ \ \\
       \vspace{-1.7cm}
       \hfill SAGA-HE-191-02 \\
                           \ \\
                           \ \\
        Modified Paschos-Wolfenstein relation
        and extraction of weak mixing angle $\bf sin^2 \theta_W$}
\author{S. Kumano}
\homepage{http://hs.phys.saga-u.ac.jp}
\email{kumanos@cc.saga-u.ac.jp}
\affiliation{Department of Physics, Saga University \\
         Saga, 840-8502, Japan}
\date{September 18, 2002}

\begin{abstract}
\vspace{0.2cm}
The NuTeV collaboration reported anomalously large weak mixing
angle $sin^2 \theta_W$ in comparison with the standard model prediction.
Neutrino and antineutrino charged- and neutral-current events are
analyzed for extracting $sin^2 \theta_W$. Although the Paschos-Wolfenstein
relation is not directly used in the analysis, it plays an important role
in the determination.
Noting that the target nucleus, iron, is not an isoscalar nucleus,
we derive a leading-order expression for a modified Paschos-Wolfenstein
relation for nuclei, which may have neutron excess.
Then, using charge and baryon-number conservations for nuclei,
we discuss a nuclear correction in the $sin^2 \theta_W$
determination. It is noteworthy that nuclear modifications are different
between valence up- and down-quark distributions.
We show this difference effect on the NuTeV $sin^2 \theta_W$ deviation.
\end{abstract}
\maketitle

\section{Introduction}\label{intro}
\setcounter{equation}{0}

The weak mixing angle $sin^2 \theta_W$ is one of the important quantities
in the standard model. In the on-shell scheme, it is related to
the $W$ and $Z$ masses by $sin^2 \theta_W = 1-m_W^2/m_Z^2$.
Collider experiments provide accurate values for these masses and
the angle. According to a global analysis \cite{lep01}, it is 
$sin^2 \theta_W^{on-shell}= 0.2227 \pm 0.0004$ 
by excluding neutrino-nucleus scattering data.

The NuTeV collaboration (G. P. Zeller, K. S. McFarland
{\it et. al.}) reported recently \cite{nutev02} 
that the mixing angle should be significantly larger: 
\begin{equation}
sin^2 \theta_W = 0.2277 \pm 0.0013 \, \text{(stat)} 
                         \pm 0.0009 \, \text{(syst)}
\, ,
\end{equation}
by using their neutrino and antineutrino scattering data.
For extracting $sin^2 \theta_W$, it is known that
the Paschos-Wolfenstein (PW) relation \cite{pw}
\begin{equation}
R^-  = \frac{  \sigma_{NC}^{\nu N}  - \sigma_{NC}^{\bar\nu N} }
              {   \sigma_{CC}^{\nu N}  - \sigma_{CC}^{\bar\nu N} }
        =  \frac{1}{2} - sin^2 \theta_W 
\, ,
\label{eqn:pw}
\end{equation}
is useful because uncertainties from charm production and 
possible nuclear effects are much reduced. 
Here, $\sigma_{NC}^{\nu N}$ and $\sigma_{CC}^{\nu N}$ are the deep inelastic
cross sections for neutral-current (NC) and charged-current (CC) neutrino
interactions with the nucleon. The factor $\rho$, which is
the relative strength between the neutral and charged currents, is taken
as one. The NuTeV collaboration measured charged and neutral current ratios,
$R_{\nu}     = \sigma_{NC}^{\nu N}/\sigma_{CC}^{\nu N}$ and
$R_{\bar\nu} = \sigma_{NC}^{\bar\nu N}/\sigma_{CC}^{\bar\nu N}$
and then a Monte Carlo simulation is used for relating the data to
$sin^2 \theta_W$. Fitting these ratios simultaneously, they end up
using the PW-like relation although it is not directly employed
in the analysis \cite{zeller}. In this sense, it is mentioned in
Ref. \cite{nutev2} that ``the NuTeV result derives $sin^2 \theta_W$
from the Paschos-Wolfenstein".
The result suggests that the left-handed neutral current coupling should
be smaller than expected. If it is confirmed, it should lead to a new physics
finding \cite{new}. The situation is summarized in the paper by
S. Davidson {\it et. al.} \cite{new}.

On the other hand, there are suggestions from a conservative point 
of view. Miller and Thomas commented \cite{mt}
that the anomalous result could be explained by the shadowing difference
between neutral and charged current reactions by using a vector meson
dominance (VMD) model. The NuTeV collaboration replied to their comments
\cite{nutev2} that the explanation is not favored because the shadowing
effects are subtracted out in the PW relation.
Furthermore, the model cannot explain observed $R_{\nu}$ and $R_{\bar\nu}$
ratios, which are smaller than expected, and also the VMD model
does not have proper $Q^2$ dependence. However, their $Q^2$
discussion is refuted by Melnitchouk and Thomas in
Ref. \cite{wally02}. Nuclear corrections are also discussed by
Kovalenko, Schmidt, and Yang \cite{ksy} by noting nuclear modifications
of $F_2$. However, one should note that such nuclear effects were
taken into account in the NuTeV analysis in a slightly different way
\cite{nutevmod}.

It is not the purpose of this paper to examine the details of these
previous studies. We rather try to address ourselves
to the extraction of $sin^2 \theta_W$ from nuclear data in a model
independent way as much as possible by resorting to charge and
baryon-number conservations. Because the NuTeV target is mainly
the iron nucleus, nuclear corrections should be carefully taken into
account for a precise determination of $sin^2 \theta_W$.
In this paper, we derive a modified PW relation for general nuclear
targets. Then, we discuss a possible nuclear modification factor
which could change the extracted $sin^2 \theta_W$ value.

This paper consists of the following. First, nuclear corrections
of the PW relation are discussed in Sec.$\,$\ref{pw}. Then,
possible effects on the extraction of $sin^2 \theta_W$
are explained in Sec.$\,$\ref{sint}. The results are summarized
in Sec.$\,$\ref{sum}.

\section{Modified Paschos-Wolfenstein relation for nuclei}\label{pw}
\setcounter{equation}{0}

The PW relation was derived for the isoscalar nucleon; however,
the used NuTeV target is mainly iron, which is not an isoscalar nucleus. 
The neutron excess may cause unexpected nuclear corrections, which
should be carefully investigated. 
In this section, we derive a leading-order (LO) PW expression
for general nuclei in a model independent way. 

First, neutrino- and antineutrino-nucleus charged current cross
sections are given in the LO by \cite{nucross}
\begin{align}
\frac{d \sigma_{CC}^{\nu A}}{dx \, dy}  =
     \sigma_{0} \,  x \, & [ \, d^A(x) +s^A(x) 
 + \{ \bar u^A(x)+ \bar c^A(x) \} (1-y)^2 \, ]
\, , 
\nonumber \\
\frac{d \sigma_{CC}^{\bar\nu A}}{dx \, dy} =
     \sigma_{0} \, x \, & [ \, \bar d^A(x)+ \bar s^A(x) 
 + \{ u^A(x)+ c^A(x) \} (1-y)^2 \, ]
\, .
\label{eqn:nucc}
\end{align}
Here, $\sigma_{0}$ is defined as $\sigma_{0}=G_F^2 s/\pi$,
by neglecting the factor $Q^2/M_W^2$ from the propagator,
with the Fermi coupling constant $G_F$ and the center-of-mass
squared energy $s$. The variables $x$ and $y$ are defined 
by the momentum transfer square $Q^2$ ($=-q^2$), the energy
transfer $q^0$, and the nucleon mass $M$ as $x=Q^2/(2 M q^0)$
and $y=q^0/E_\nu$ or $q^0/E_{\bar\nu}$. 
Nuclear quark and antiquark distributions are denoted 
$q^A(x)$ and $\bar q^A(x)$, respectively. 
They depend also on $Q^2$; however, the explicit $Q^2$ dependence
is abbreviated in the parton distribution functions (PDFs) throughout
this paper. Next, the neutral current cross section
for neutrino scattering is given by
\cite{nucross}
\begin{align}
\frac{d \sigma_{NC}^{\nu A}}{dx \, dy}  = &
      \sigma_{0} \, x \, [ \, 
   \{ u_L^2 + u_R^2 (1-y)^2 \} \{ u^A(x) + c^A(x) \} 
\nonumber \\        
&  + \{ u_R^2 + u_L^2 (1-y)^2 \} \{ \bar u^A(x) + \bar c^A(x) \} 
\nonumber \\ 
&  + \{ d_L^2 + d_R^2 (1-y)^2 \} \{ d^A(x) + s^A(x) \} 
\nonumber \\        
&  + \{ d_R^2 + d_L^2 (1-y)^2 \} \{ \bar d^A(x) + \bar s^A(x) \} 
\, ]
\, ,
\label{eqn:nunc}
\end{align}
where left- and right-handed couplings are expressed by
the weak mixing angle as
$u_L = 1/2- (2/3) \, sin^2 \theta_W$,
$u_R = -(2/3) \, sin^2 \theta_W$,
$d_L = -1/2 +(1/3) \, sin^2 \theta_W$, and
$d_R = (1/3) \, sin^2 \theta_W$.
The cross section for antineutrino scattering is obtained by
exchanging the left- and right-handed couplings.

Using these equations, we obtain a nuclear PW relation as
\begin{align}
R_A^- & = \frac{   \sigma_{NC}^{\nu A} - \sigma_{NC}^{\bar\nu A} }
                  {   \sigma_{CC}^{\nu A} - \sigma_{CC}^{\bar\nu A} }
\nonumber \\
       & \! \! \! \! \! \! \! \! \! \! 
 = \{ 1-(1-y)^2 \} \,  [ \, (u_L^2 -u_R^2 ) \{ u_v^A(x) + c_v^A (x) \}
\nonumber \\
       &  \ \ \ \ \ \ \ \ \ \ \ \ \ \ \ \ 
+  (d_L^2 -d_R^2 ) \{ d_v^A (x) + s_v^A (x) \} \, ]
\nonumber \\
       &  \! \! \! \! \! \! \! \!
 / \, [ \, d_v^A (x) + s_v^A (x) 
         - (1-y)^2 \, \{ u_v^A (x) + c_v^A (x) \} \, ]
\, .
\label{eqn:apw1}
\end{align}
Here, the valence quark distributions are defined by
$q_v^A \equiv q^A -\bar q^A$.
The name , valence quark, is conventionally used for valence up- and
down-quark distributions, but we use the same nomenclature for strange
and charm distributions by applying the same definitions,
$s_v^A \equiv s^A  -\bar s^A $
and $c_v^A \equiv c^A -\bar c^A $.
Of course, there is no net strangeness and charm in ordinary nuclei,
so that ``valence" strange and charm distributions should satisfy
$\int dx \, s_v^A (x)=0$ and $\int dx \, c_v^A (x)=0$.
However, these equations do not mean that the distributions themselves
vanish: $s_v^A (x) \ne 0$ and $c_v^A (x) \ne 0$. 

The valence-quark distributions $u_v^A$ and $d_v^A$ are expressed by
the weight functions $w_{u_v}$ and $w_{d_v}$ at any $Q^2$:
\begin{align}
u_v^A (x) & = w_{u_v} (x,A,Z) \, \frac{Z \, u_v (x) + N \, d_v (x)}{A},
\nonumber \\
d_v^A (x) & = w_{d_v} (x,A,Z) \, \frac{Z \, d_v (x) + N \, u_v (x)}{A},
\label{eqn:wpart}
\end{align}
where $u_v$ and $d_v$ are the distributions in the proton,
$Z$ is the atomic number, $N$ is the neutron number, and
$A$ is the mass number of a nucleus.
The weight functions are defined at fixed $Q^2$ (=1 GeV$^2$)
in Ref. \cite{saga01}; however, they are used at any $Q^2$ 
throughout this paper.
The explicit $Q^2$ dependence is abbreviated as for the PDFs.
Although these equations are originally motivated
by the isospin symmetry in nuclei for the virtual $w_{u_v}=w_{d_v}=1$
case, we do not rely on such an assumption.
This is because nuclear modifications, including isospin violation
\cite{isospin,sv} and nuclear charge-symmetry breaking \cite{charge},
could be taken into account by the weight functions in any case.
Therefore, the expressions are given without losing any generality.

Next, we define neutron excess constant $\hat\varepsilon_n$ and
a related function $\varepsilon_n (x)$ by
\begin{equation}
\hat\varepsilon_n =\frac{N-Z}{A} \, , \ \ \ 
\varepsilon_n (x) = \hat\varepsilon_n 
            \frac{u_v(x)-d_v(x)}{u_v(x)+d_v(x)} \, .
\label{eqn:en}
\end{equation}
Then, a difference between the weight functions is defined by
\begin{equation}
\varepsilon_v (x) = \frac{w_{d_v}(x,A,Z)-w_{u_v}(x,A,Z)}
                             {w_{d_v}(x,A,Z)+w_{u_v}(x,A,Z)}
\, .
\label{eqn:ev}
\end{equation}
The function $\varepsilon_v$ depends also on $A$, $Z$, and $Q^2$,
but these factors are abbreviated in writing $\varepsilon_v$.
Substituting Eqs. (\ref{eqn:wpart}), (\ref{eqn:en}),
and (\ref{eqn:ev}) together with the coupling constants into
Eq. (\ref{eqn:apw1}), we obtain
\begin{align}
& R_A^-   =  \bigg [ \,
\bigg ( \frac{1}{2} - sin^2 \theta_W \bigg ) \,
\{ 1+ \varepsilon_v (x) \, \varepsilon_n (x) \}
\nonumber \\
       &  \ 
+\frac{1}{3} \, sin^2 \theta_W \, 
\{ \varepsilon_v (x)  + \varepsilon_n (x) \}
+ \bigg ( \frac{1}{2} - \frac{2}{3} \, sin^2 \theta_W \bigg ) \, 
  \varepsilon_s (x)
\nonumber \\
       &  \ 
+ \bigg ( \frac{1}{2} - \frac{4}{3} \, sin^2 \theta_W \bigg ) \,
  \varepsilon_c (x) \bigg ] \, 
\bigg /  \, \bigg [ \,
1+ \varepsilon_v (x) \, \varepsilon_n (x)
\nonumber \\ 
       &   \! \! \! \! \! \! 
+ \frac{1+(1-y)^2}{1-(1-y)^2} 
 \{ \varepsilon_v (x) + \varepsilon_n (x)  \}
+\frac{2 \{ \varepsilon_s (x) - (1-y)^2 \varepsilon_c (x) \}}{1-(1-y)^2} 
\, \bigg ]
\, .
\label{eqn:apw2}
\end{align}
Here, $\varepsilon_s$ and $\varepsilon_c$ are defined by
$\varepsilon_s = s_v^A /[w_v \, (u_v+d_v)]$ and
$\varepsilon_c = c_v^A /[w_v \, (u_v+d_v)]$
with $w_v = (w_{d_v}+w_{u_v})/2$.
We would like to stress that the LO expression in Eq. (\ref{eqn:apw2})
has been derived without using any model dependent factor and
and any serious approximation.

The strange quark ($s_v^A$) effects are investigated in Ref. \cite{sv},
and the neutron-excess effects, namely the $\varepsilon_n$ terms, are
taken into account in the NuTeV analysis \cite{nutevmod}.
In addition to these corrections, we notice in Eq. (\ref{eqn:apw2})
that another correction factor $\varepsilon_v$ contributes to the deviation
from the PW relation $1/2 - sin^2 \theta_W$ due to
{\it the difference between the valence up- and
     down-quark modifications in a nucleus.}
This fact needs to be clarified.
There are two restrictions on the valence-quark distributions in a nucleus.
One is the baryon number conservation \cite{fsl}, and 
the other is the charge conservation \cite{saga01}.
Nuclear baryon number and charge have to be $A$ and $Z$, and they
are expressed in the parton model as
\begin{align}
&  \! \! \! 
A  = \! \!  \int dx \, A   \sum_q \frac{1}{3} \, (q^A - \bar q^A)
   = \! \!  \int dx  \, \frac{A}{3} \, ( u_v^A + d_v^A )
\, , \nonumber \\
&  \! \! \! 
Z  = \! \!  \int dx \, A   \sum_q e_q  (q^A - \bar q^A)
   = \! \!  \int dx \, \frac{A}{3} \, ( 2 \, u_v^A - d_v^A )
\, ,
\label{eqn:conserv}
\end{align}
where $A$ is multiplied in the integrands because the nuclear parton
distributions are defined by those per nucleon, and the relation
$\int dx \, s_v (x) =\int dx \, c_v (x) =0$ is used in obtaining
the right-hand sides. 
Substituting Eq. (\ref{eqn:wpart}) into Eq. (\ref{eqn:conserv}), we obtain
\begin{align}
\int dx \, (u_v+d_v) \, & [ \, \Delta w_v 
         +  w_v \, \varepsilon_v (x) \, \varepsilon_n (x) \, ] = 0
\, ,
\label{eqn:b} \\
\int dx \, (u_v+d_v) \, & [ \, \Delta w_v \, 
         \{ 1-3 \, \varepsilon_n(x) \} \,
\nonumber \\
&        - w_v \, \varepsilon_v (x) \, 
         \{ 3 - \varepsilon_n (x) \}  \, ] =0
\, ,
\label{eqn:c}
\end{align}
where $\Delta w_v$ is defined by $\Delta w_v=w_v -1$.
Although it is not straightforward to determine $\varepsilon_v (x)$ 
from Eqs. (\ref{eqn:b}) and (\ref{eqn:c}), it indicates that
$\varepsilon_v (x)$ is finite.

In this way, we find that nuclear modifications
are, in general, different between the valence up- and down-quark
distributions because of the baryon number and charge conservations.
It gives rise to the factor $\varepsilon_v$.
This kind of detailed nuclear effects cannot be accounted simply by
investigating electron and muon scattering data and 
by fitting charged current cross-section data for the same target
\cite{nutevmod}. Because the physics associated with the $\varepsilon_v$
factor is missing in the NuTeV analysis, it may cause a significant effect
on the $sin^2 \theta_W$ determination.

\section{Effects on $\bf sin^2 \theta_W$ determination}\label{sint}
\setcounter{equation}{0}

The angle $sin^2 \theta_W$ can be extracted by using Eq. (\ref{eqn:apw2})
together with the experimental data of $R_A^-$. In order to find whether
or not the $\varepsilon_n$ corrections could explain the deviation from the
standard model value for $sin^2 \theta_W$, we approximate 
the relation by considering that the corrections are small
($\ \varepsilon_v \ll 1$).
Retaining only the leading correction of $\ \varepsilon_v$
in Eq. (\ref{eqn:apw2}), we obtain
\begin{align}
& R_A^-   =  \frac{1}{2} - sin^2 \theta_W  
\nonumber \\
       &  \ 
- \varepsilon_v (x)  \bigg \{ \bigg ( \frac{1}{2} - sin^2 \theta_W \bigg ) 
               \frac{1+(1-y)^2}{1-(1-y)^2} - \frac{1}{3} sin^2 \theta_W 
\bigg \}
\nonumber \\
       &  \ 
+O(\varepsilon_v^2)+O(\varepsilon_n)+O(\varepsilon_s)+O(\varepsilon_c) 
\, .
\label{eqn:apw3}
\end{align}
The higher-order and other corrections $O(\varepsilon_v^2)$,
$O(\varepsilon_n)$, $O(\varepsilon_s)$, and $O(\varepsilon_c)$ are not
explicitly written.
As far as present neutrino data suggest, the ``valence"
strange distribution should be small, and the measurements indicate
that such a correction increases the NuTeV $sin^2 \theta_W$ deviation
according to Ref. \cite{sv}.
Therefore, at least at this stage, the finite $s_v^A$ and $c_v^A$
distributions effects, $O(\varepsilon_s)$ and $O(\varepsilon_c)$, 
are not the favorable explanation.
Although accurate measurements may clarify the details of
the distributions $s_v^A$ and $c_v^A$ in future, they are not
discussed in the following.
The neutron-excess effects $O(\varepsilon_n)$ are included in
the NuTeV analysis \cite{nutevmod}, so that they are not
the source of the $sin^2 \theta_W$ deviation.

As mentioned in Sec. \ref{intro} and Ref. \cite{nutev2},
the NuTeV $sin^2 \theta_W$ is ``derived" from the PW-like relation
indirectly. It is obvious from Eq. (\ref{eqn:apw3}) that there is
an $\varepsilon_n$-type correction to $sin^2 \theta_W$, and it may explain,
at least partially, the deviation from the standard model.
However, the correction is essentially unknown at this stage 
in the sense that there is no significant data to find the difference
between valence up- and down-quark modifications ($\varepsilon_v$).
In order to investigate whether or not the $\varepsilon_v$ correction
is large enough, we should inevitably use some prescription for
describing the $\varepsilon_v$ factor.
In the following, we introduce two different descriptions as examples.

\begin{center}
{\bf 1. A prescription for the conservations}
\end{center}

It is not straightforward to find a solution $\varepsilon_v(x)$
to satisfy Eqs. (\ref{eqn:b}) and (\ref{eqn:c}). For an approximate
estimate, the higher-order corrections $\varepsilon_v \varepsilon_n$
are neglected in these equations. Then, substituting Eq. (\ref{eqn:b})
into Eq. (\ref{eqn:c}), we obtain
\begin{equation}
\varepsilon_v (x) 
= - \hat \varepsilon_n \, 
               \frac{u_v(x)-d_v(x)}{u_v(x)+d_v(x)} \,
               \frac{\Delta w_v(x)}{w_v(x)}
\, ,
\label{eqn:evx2}
\end{equation}
by considering a special case that the integrand vanishes.
Of course, this is not a unique solution, but this estimate should be able
to provide information about the magnitude of the correction.

\begin{center}
{\bf 2. A $\bf\chi^2$ analysis of nuclear PDFs}
\end{center}

Global $\chi^2$ analysis results could be used for calculating
$\varepsilon_v (x)$. A $\chi^2$ analysis for determining nuclear parton
distribution functions (NPDFs) is reported in Ref. \cite{saga01}, and
obtained distributions can be calculated by using subroutines at the web site
in Ref. \cite{sagaweb}. Before using the NPDF code, we would like to remind
the reader what has been done for the valence distributions.
At $Q^2$=1 GeV$^2$, the weight functions  are assumed to be
\begin{equation}
\! \! \! \!
w_{i}(x)  = 1+\left( 1 - \frac{1}{A^{1/3}} \right)
         \frac{a_{i} +b_v x+c_v x^2 +d_v x^3}{(1-x)^{\beta_v}},
\label{eqn:wudv}
\end{equation}
where $i$ denotes $u_v$ or $d_v$, and
$a_{u_v}$, $b_v$, $c_v$, and $d_v$ are the parameters to be
determined by the $\chi^2$ fit. Because it is the first $\chi^2$ analysis
attempt for nuclei, the parameter number has to be reduced as many as possible.
It is the reason why the common parameters are chosen for $\beta_v$,
$b_v$, $c_v$, and $d_v$. However, in order to satisfy both the charge
and baryon-number conservations, at least one parameter should be different.
Because the parameters $a_{u_v}$ and $a_{d_v}$ are determined
so as to satisfy these conservations, this description of
$\varepsilon_v (x)$ surely satisfies Eqs. (\ref{eqn:b}) and (\ref{eqn:c}). 
However, one should note that this $\varepsilon_v (x)$ may not be valid
if it has more complicated $x$ dependence than the one calculated
from Eq. (\ref{eqn:wudv}).

\subsection{Numerical results}\label{results}

In the first description, $\varepsilon_v$ is evaluated numerically
by using Eq. (\ref{eqn:evx2}) with the NPDF code in Ref. \cite{sagaweb}
for calculating $u_v$, $d_v$, $w_{u_v}$, and $w_{d_v}$ at given $Q^2$.
In the second one, $\varepsilon_v$ is calculated by the definition
in Eq. (\ref{eqn:ev}) with the weight functions $w_{u_v}$ and $w_{d_v}$,
which are numerically calculated by the NPDF code.
In the NuTeV measurements \cite{zeller,nutev2}, averages of
the kinematical variables are given by 
$<E>$=120 and 112 GeV, 
$<Q^2>$=25.6 and 15.4 GeV$^2$,
$<x>$=0.22 and 0.18 for neutrinos and antineutrinos, respectively.
The ranges of  $x$ and $Q^2$ are $0.01 < x < 0.75$
and $1 < Q^2 < 140 \text{ GeV$^2$}$.
Although the $\nu$ and $\bar\nu$ incident energies are different,
we use the averaged value $<E>$=116 GeV
for connecting $x$, $y$, and $Q^2$:
$y=Q^2/(2 M x <E>)$.

\begin{figure}[h]
\includegraphics[width=0.40\textwidth]{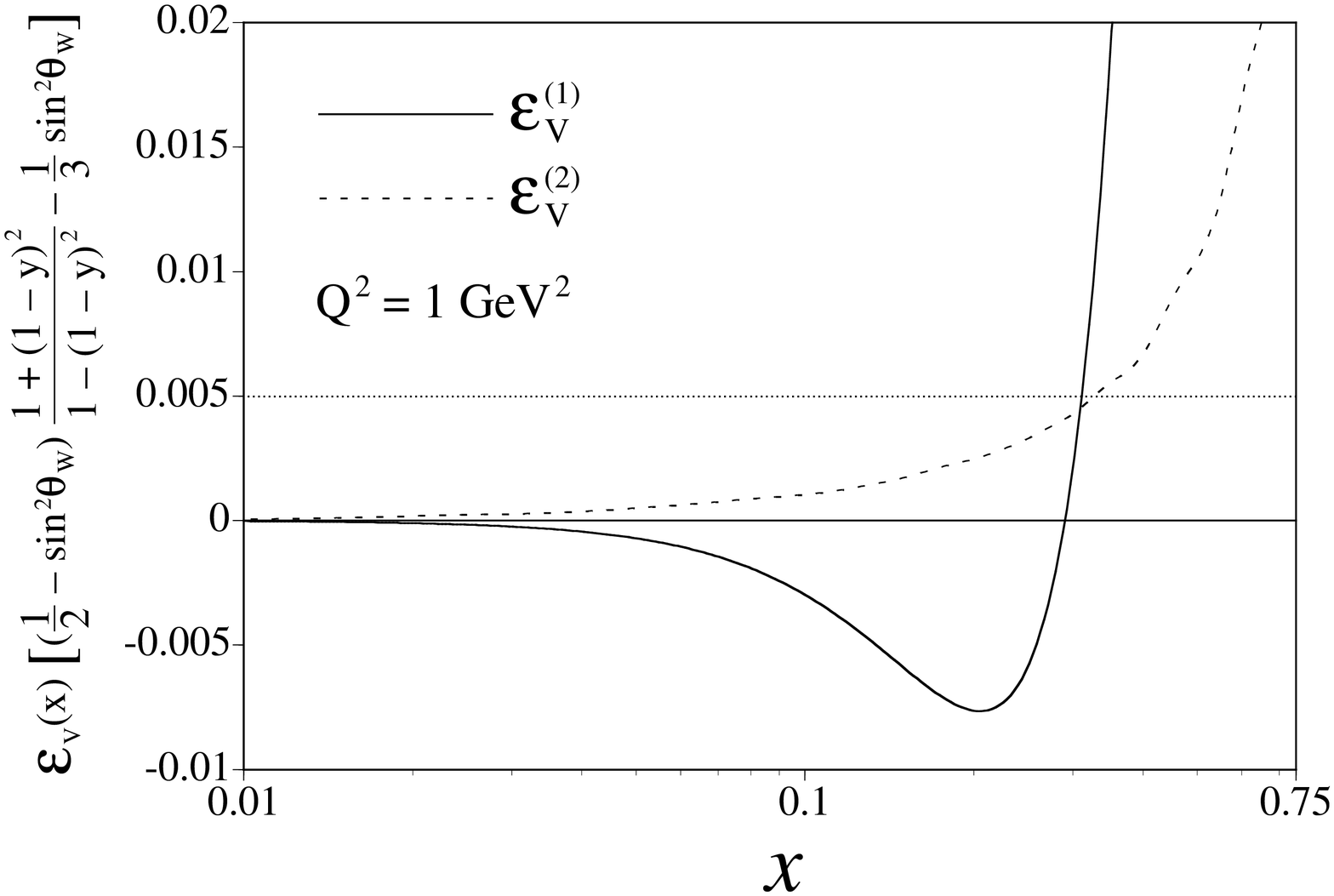}
\vspace{-0.5cm}
\caption{The $\varepsilon_v$ correction term is evaluated 
         at $Q^2$=1 GeV$^2$. The solid curve is calculated by 
         Eq. (\ref{eqn:evx2}), and the dashed curve is obtained by
         the $\chi^2$ fit code \cite{saga01,sagaweb}. 
         The NuTeV deviation 0.005 is shown by the dotted line
         just for comparison.}
\label{fig:xdep1}
\end{figure}

In Fig. \ref{fig:xdep1}, the $Q^2$ value is fixed at $Q^2=1$ GeV$^2$,
and the correction term in Eq. (\ref{eqn:apw3}) is evaluated
as a function of $x$. The neutron excess $\hat\varepsilon_n=4/56$
and $sin^2 \theta_W= 0.2227$ are used in the calculations.
The solid ($\varepsilon_v^{(1)}$) and dashed ($\varepsilon_v^{(2)}$)
curves are obtained by the first and second descriptions, respectively.
The dotted line indicates the NuTeV $sin^2 \theta_W$ deviation
(0.2277$-$0.2227=0.0050) just for a comparison.
Both curves increase rapidly as $x$ becomes larger, and this is
mostly a kinematical effect due to the factor $1/[1-(1-y)^2]$.
Another effect at large $x$ comes from the Fermi-motion-like factor
$1/(1-x)^{\beta_v}$ in Eq. (\ref{eqn:wudv}).
The $\varepsilon_v^{(2)}$ correction term is very small with
the following reason. The $x$ dependence in Eq. (\ref{eqn:wudv})
indicates that $\varepsilon_v^{(2)}$ is almost independent of $x$
except for large $x$.
The first terms in Eqs. (\ref{eqn:b}) and (\ref{eqn:c}) are
valence-quark distributions multiplied by $\Delta w_v$,
and these integrals almost vanish. The obtained $\varepsilon_v^{(2)}$
is roughly proportional to these small integral values.
One should note that this result could be artificially small
due to the weak $x$ dependence assumption.
On the other hand, the magnitude of $\varepsilon_v^{(1)}$ is
large in general and it has completely different $x$ dependence.
Because $\varepsilon_v^{(1)}$ is directly proportional to
the valence-quark modification ($\Delta w_v$) in Eq. (\ref{eqn:evx2}),
the function changes sign at $x \approx 0.3$, which is
the transition point from antishadowing to the EMC-effect region.
In the revised analysis \cite{saga02,skf3}, the valence
modifications are slightly different; however, the essential features
are the same. According to the first description, the function
becomes comparable magnitude to the NuTeV deviation.

\begin{figure}[h]
\includegraphics[width=0.40\textwidth]{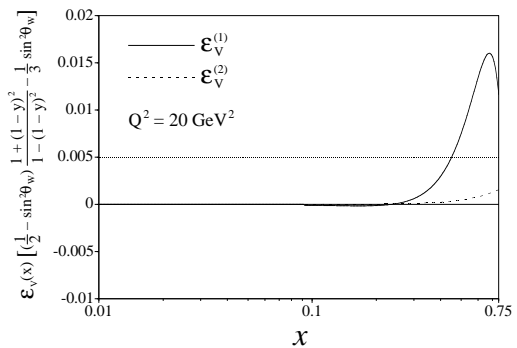}
\vspace{-0.5cm}
\caption{The correction term is evaluated at $Q^2$=20 GeV$^2$.}
\label{fig:xdep20}
\end{figure}

Because the average $Q^2$ is much larger than 1 GeV$^2$ in the NuTeV
experiment, the corrections are also calculated at $Q^2$=20 GeV$^2$ by
noting the kinematical limit $y<1$. 
The results are shown in Fig. \ref{fig:xdep20}. 
In comparison with the $Q^2$=1 GeV$^2$ results, the effects are much
suppressed. This is again due to the kinematical factor $1/[1-(1-y)^2]$. 
For example, this factor is 55 for $x$=0.5 and $Q^2$=1 GeV$^2$;
however, it becomes 3.0 for $x$=0.5 and $Q^2$=20 GeV$^2$.
This is the reason why both distributions become smaller.
Although the $\varepsilon_v^{(2)}$ effect is too small to explain
the deviation at $Q^2$=20 GeV$^2$, the $\varepsilon_v^{(1)}$ 
is still comparable magnitude. 

\begin{figure}[h]
\includegraphics[width=0.40\textwidth]{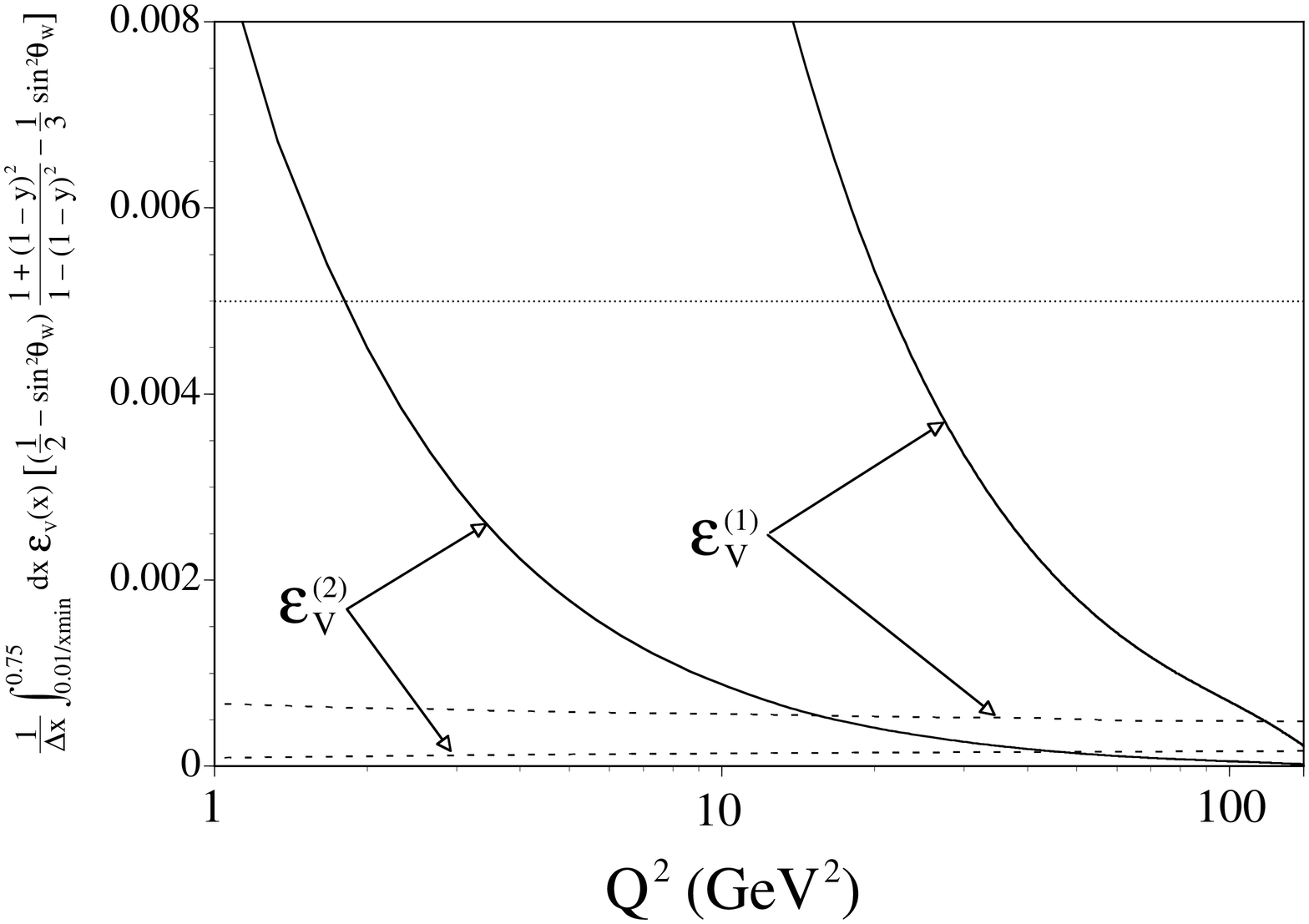}
\vspace{-0.5cm}
\caption{The solid curves indicate
       the corrections averaged over the $x$ range. 
       The dashed ones are obtained by taking the NuTeV kinematics 
       into account.}
\label{fig:q2dep}
\end{figure}

In oder to investigate these effects on the NuTeV $sin^2 \theta_W$,
the analysis should be done with the Monte Carlo code
with the experimental data. However, in order to show the order of
magnitude, we simply average the obtained curves
over the $x$ range ($\Delta x$) from 0.01 or $x_{min}=Q^2/(2M<E>)$ to 0.75.
Because the data are centered at about $x=0.2$, 
this kind of simple average could overestimate
the effects coming from the large $x$ region. 
The calculated results are shown by the solid curves
in Fig. \ref{fig:q2dep}. 
As already found in Figs. \ref{fig:xdep1} and \ref{fig:xdep20},
the effects are very large at small $Q^2$ and become smaller
as $Q^2$ becomes larger. 

If the simple $x$ average is taken, the effects look large.
On the other hand, there is a method to take the NuTeV kinematics
into account \cite{kevin} by using the functionals in Fig. 1
of Ref. \cite{sv}. Although the physics motivation is completely
different, the present $\varepsilon_v$ distribution could be simulated
by the NuTeV distributions $u_v^p - d_v^n$ and $d_v^p - u_v^n$.
It is interesting that their ``isospin-violating distributions"
could effectively contain the present nuclear effect.
If such a correspondence is made, the effect on $sin^2 \theta_W$
is calculated by using the NuTeV functionals \cite{zeller,kevin}.
The results are shown by the dashed curves in Fig. \ref{fig:q2dep}.
We find that the effects become significantly smaller due to
the lack of large $x$ events in the NuTeV experiment.
Therefore, as far as the considered two descriptions are
concerned, the nuclear modifications
are not large enough to explain the whole NuTeV deviation.

It is, however, still too early to conclude that the present
mechanism is completely excluded, because the nuclear modification
difference between $u_v^A$ and $d_v^A$ is not measured at all.
It may not be possible to find this nuclear effect until
a neutrino factory \cite{skf3,nufact} or the NuMI project \cite{numi}
is realized. On the other hand, if these facilities are built,
the ``nucleon" cross sections (hence $sin^2 \theta_W$) could be measured
with proton and deuteron targets with minor nuclear corrections.
The NuTeV also reported that $R_\nu$ and
$R_{\bar\nu}$ are unexpectedly smaller \cite{nutev2}.
We are also investigating this issue, and hopefully it will be clarified
in the near future. 

\vspace{0.8cm}
\section{Summary}\label{sum}

We have derived a modified Paschos-Wolfenstein relation for nuclei.
Using this relation, we investigated the possibility that
the NuTeV $sin^2 \theta_W$ deviation could be explained by the nuclear
parton distributions in iron. In particular, we pointed out that nuclear
modifications are different between valence up- and down-quark distributions.
The difference partially explains the NuTeV deviation although
it may not be large enough to explain the whole deviation.
Because such modifications are not measured, it is important
to investigate them in future.



\end{document}